\documentclass[conference]{IEEEtran}
\IEEEoverridecommandlockouts
\usepackage{cite}
\usepackage{amsmath,amssymb,amsfonts}
\usepackage{algorithmic}
\usepackage{graphicx}
\usepackage{textcomp}
\usepackage{xcolor}
\usepackage{CJK}
\usepackage{url}
\usepackage{booktabs}
\usepackage{caption}
\usepackage{makecell}
\usepackage{multirow}
\usepackage{soul}
\usepackage{hyperref}

\usepackage{amsmath,amssymb,inputenc,graphicx,xcolor,rotating,amsthm}
\usepackage{blindtext}
\usepackage{enumerate}
\usepackage{xcolor}
\usepackage{tcolorbox}
\tcbset{center, top=5pt, bottom=5pt, width = \textwidth,colback=pink!65!white,colframe=pink!10!}

\newcommand{\RR}{\mathbb{R}}
\newcommand{\CC}{\mathbb{C}}

\definecolor{darkviolet}{rgb}{0.58,0,0.83} 

\begin{document}


\title{ISAC: An Invertible and Stable Auditory Filter Bank with
Customizable Kernels for ML Integration\\
\thanks{D. Haider and F. Perfler are recipients of a DOC Fellowship (A 26355 and A 27288, respectively) of the Austrian Academy of Sciences at the Acoustics Research Institute. The work of N. Holighaus was supported by the FWF project DISCO (PAT4780023). 
The work of P. Balazs was supported by the FWF projects LoFT (P 34624), NoMASP (P 34922), Voice Prints (P 36446) and the WWTF project EleCom (LS23-024). 
C. Hollomey has been funded by the Vienna Science and Technology Fund
(WWTF) and by the State of Lower Austria [Grant ID: 10.47379/LS23024]. ISAC was derived from the implementation of \texttt{audfilters} of the Large Time-Frequency Analysis Toolbox (LTFAT) (https://ltfat.org).}
}

\author{
\IEEEauthorblockN{Daniel Haider\textsuperscript{*}, Felix Perfler\textsuperscript{*}, Peter Balazs\textsuperscript{*}\IEEEauthorrefmark{3}, Clara Hollomey\textsuperscript{$\dagger$}, and Nicki Holighaus\textsuperscript{*}}
\IEEEauthorblockA{\textit{\textsuperscript{*}Acoustics Research Institute, Austrian Academy of Sciences}, Vienna, Austria \\
\textit{\textsuperscript{$\dagger$}University of Applied Sciences St. Pölten}, St. Pölten, Austria \\
\IEEEauthorblockA{\IEEEauthorrefmark{3} \textit{Interdisciplinary Transformation University Austria (IT:U)}, Linz, Austria}
daniel.haider@oeaw.ac.at}
}

\maketitle

\begin{abstract}
    This paper introduces ISAC, an invertible and stable, perceptually-motivated filter bank that is specifically designed to be integrated into machine learning paradigms. More precisely, the center frequencies and bandwidths of the filters are chosen to follow a non-linear, auditory frequency scale, the filter kernels  have user-defined maximum temporal support and may serve  as learnable convolutional kernels, and there exists a corresponding filter bank such that both form a perfect reconstruction pair. ISAC  provides a powerful and user-friendly audio front-end suitable for any application, including analysis-synthesis schemes.
\end{abstract}


\section{Introduction}


Using learnable time-frequency filter banks as front-ends for neural networks can considerably increase the computational efficiency of a wide range of audio processing tasks~\cite{nnaudio}. Related approaches range from fully learnable convolutional layers with 1D kernels~\cite{sainath2013fblearing, baevski2020wav2vec, lou2019convtasnet} to parametrized constructions where only the bandwidths and center frequencies of fixed prototype kernels are learned~\cite{zeghidour2021leaf, ravanelli2018sincnet}. Prior work indicates that optimizing learnable filter bank front-ends benefits from initialization as perceptually-motivated time-frequency filters, e.g., based on the mel frequency scale~\cite{zeghidour2018learning}, and a wide range of related transforms has been used~\cite{fu2022fastaudio, sailor2017unsupervised, yu2017dnn}.

Although invertibility and numerical stability of the underlying time-frequency filter bank is of limited concern for classification tasks, it is crucial in analysis-synthesis schemes~\cite{lou2019convtasnet, braun2020data, haider2024hold} to prevent the introduction of errors and ensure that no needed information is lost. Moreover, it has been found that stability in the form of a 1-Lipschitz or even Parseval property significantly increases robustness of the model against noise and adversarial examples~\cite{nenov2024smoothsailing, balazs2024trainable, cisse2017parseval, hasannasab2020proximal}.

Perceptually-motivated filter banks that are perfectly invertible and numerically stabile exist~\cite{necciari2018AUDlet, derrien2015quasi}. In particular, 
these constructions employ filters with bandwidths matched to the \textit{critical bands}~\cite{zwicker1961subdivision} of human hearing. However, the frequency-domain implementation in~\cite{necciari2018AUDlet, derrien2015quasi} does not provide control of the kernel size\footnote{In the signal processing literature, it is common to refer to the convolution kernel as \emph{impulse response}. Here, we prefer the term convolution kernel, which is widely understood in the mathematics and machine learning communities.}, even favoring bandlimited filters, which necessarily have kernels of infinite length. Yet, control of the kernel size is a requirement for the efficient integration of any time-frequency filter bank into learnable machine learning frameworks to restrict the number of learnable parameters and employ linear convolution directly on the GPU~\cite{nnaudio, kapre}.

\begin{figure}[t]
    \centering
    \begin{minipage}{\linewidth}
        \includegraphics[width=\linewidth]{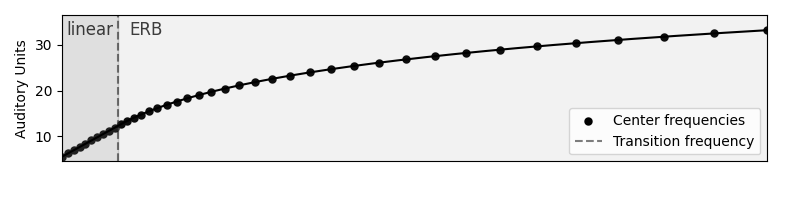}
    \end{minipage}\par
    \vspace{-10pt}  

    \begin{minipage}{\linewidth}
        \includegraphics[width=\linewidth]{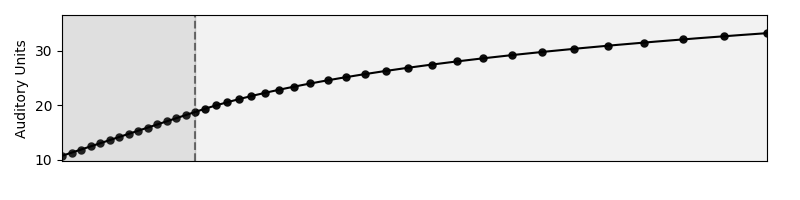}
    \end{minipage}\par
    \vspace{-10pt}  

    \begin{minipage}{\linewidth}
        \includegraphics[width=\linewidth]{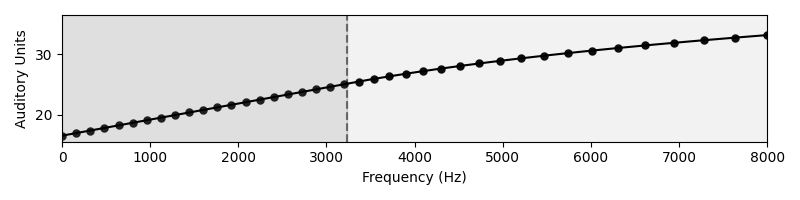}
    \end{minipage}

    \caption{Frequency scales of ISAC filter banks for kernel sizes $256$, $128$, and $64$ (top to bottom), based on the ERB scale. The center frequencies in the dark area lie on a linear scale, and those in the light area on the ERB scale. The shorter the kernels, the larger the linear part of the scale becomes.}
    \label{fig:scale}
\end{figure}

We fill this gap by proposing ISAC, a perceptually-motivated time-frequency filter bank that is invertible and numerically stable. While ISAC is based on the construction proposed in~\cite{necciari2018AUDlet}, we introduce some key changes to enhance user-friendliness and facilitate an efficient integration into convolutional machine learning frameworks. On demand, the kernels of ISAC can be further optimized together with the remaining model parameters, similar to~\cite{zeghidour2018learning}. Stability and invertibility can be maintained throughout the whole training process by including a regularization term in the learning objective~\cite{balazs2024trainable, haider2024hold, cisse2017parseval}. Moreover, we can compute structurally identical approximate duals, i.e., having the same kernel sizes.

The central novelty in ISAC is a user-defined maximal kernel size. In order to achieve this while maintaining bandwidth requirements and a flat total power spectral density (PSD),
we linearize the auditory scale below the frequency at which the desired kernel size is first achieved.
We thus obtain a set of filters with short kernels that can be used for time-frequency processing in various applications. Our implementation supports the seamless integration into machine learning frameworks and, similar to the idea in~\cite{zeghidour2021leaf}, it is also possible to extend the filter bank to a compressed mel-type spectrogram.

\begin{figure}[t]
    \centering
    \includegraphics[width=1\linewidth]{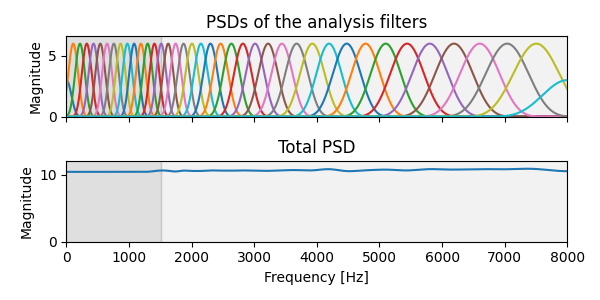}
    \caption{Individual and total power spectral densities of an ISAC filter bank with $40$ kernels of size $T^*=128$. The flat total PSD indicates that the filter bank is well conditioned. The condition number for a decimation factor of $6$ is $\kappa = 1.05$.}
    \label{fig:analysis}
\end{figure}

\section{Filter Banks and Convolutional Layers}


\emph{Filter banks} and \emph{(linear) convolutional layers} are two names for the same class of linear, time-invariant (LTI) operators, possibly combined with decimation. They are well-suited for processing data that exhibits translation invariance. In practice, however, the two terms hint at different use cases and implied restrictions to the considered LTI operator. \emph{Time-frequency transforms} are designed as filter banks to match specific time-frequency characteristics, with kernel sizes typically ranging from $10$ to about $100$ ms (corresponding to $48$ to $4800$ samples for a sampling rate $f_s=48$ kHz) for most audio processing tasks.
Learned convolutional layers, on the other hand, usually employ much shorter kernels for processing audio input. As an example, the default kernel size in the learnable encoder of ConvTasNet~\cite{lou2019convtasnet} is $16$, and wav2vec~\cite{baevski2020wav2vec} as well as WavLM~\cite{chen2022wavlm} use kernel lengths between $2$ and $10$ samples to represent the input audio.
Reasons for this comprise the decreased computational demands along with empirically observed performance improvements associated with employing shorter kernels~\cite{zhao2024review, leakage21}.
Moreover, long filters are prone to numerical instabilities~\cite{haider2023randomfb} and may increase the risk of overfitting when training data is limited. To capture dependencies over a longer time range with short kernels, the models include different dimensionality reduction schemes and are usually designed deeper, i.e., multiple convolutional layers with short kernels are composed. 
Hence, integrating a learnable filter bank into this paradigm, necessitates to allow a customization of the kernel sizes to a definable maximum, similar to how we set them for time-frequency filter banks in practice.

In mathematical terms, for a finite discrete real-valued audio signal $x \in \mathbb{R}^L$ and a filter $g\in \CC^L$ (specified by its kernel), filtering $x$ by $g$ equals finite circular convolution of $x$ and $g$ (i.e., indices are understood $\mathrm{mod}\;L$) and decimate the output by a factor $d\geq1$ (a.k.a. stride), classically written as
\begin{equation}\label{eq:conv1}
    (x \ast g)\downarrow_d[n] = \sum_{k=0}^{L-1} x[k] g[dn-k].
\end{equation}
Assuming that a filter has kernel size $T \ll L$, by commutativity of the convolution, we may rewrite the above as
\begin{equation}\label{eq:conv2}
    (x \ast g)\downarrow_d[n] = \sum_{k=0}^{T-1} g
    [k] x[dn-k].
\end{equation}
While \eqref{eq:conv1} is how convolution is formally defined, the reduced version in \eqref{eq:conv2} is how convolution (with circulant boundary conditions) is applied in a convolutional layer, where the $T$ kernel coefficients are the trainable parameters of the network.

\begin{figure}[t]
    \centering
    \includegraphics[width=1\linewidth]{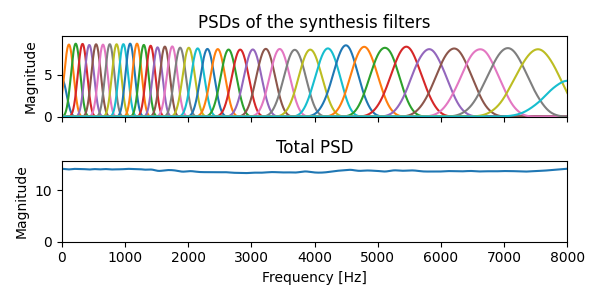}
    \caption{Individual and total power spectral densities of learned synthesis kernels for the setting of Fig. \ref{fig:analysis}. The reconstruction error is 6e-6. A regularizing term ensured that the total PSD stays flat. The condition number is $\kappa = 1.07$.}
    \label{fig:synthesis}
\end{figure}

\section{The ISAC Filter Bank Construction}
The construction of the ISAC filters closely follows the one for AUDlets~\cite{necciari2018AUDlet}. We recall the main idea: Given a (non-linear) auditory scale function $F_S:[0,\tfrac{f_s}{2}]\rightarrow S$ that maps positive frequencies (in Hz) to an auditory scale $S$ (in auditory units), and a function $B_S:[0,\tfrac{f_s}{2}]\rightarrow \RR$ that gives the associated bandwidths. For the commonly used equivalent rectangular bandwidth (ERB) auditory scale~\cite{glasberg1990derivation,Necciari:2013a}, the functions are given by
\begin{equation}\label{eq:erb}
    F_{\text{ERB}}(f) = 9.265 \ln \left( 1 + \frac{f}{228.8455} \right),
\end{equation}
and
\begin{equation}\label{eq:erbbw}
    B_{\text{ERB}}(f) = 24.7 + \frac{f}{9.265}.
\end{equation}
If there is no associated bandwidth function (e.g., for the mel scale), setting $B_S(f)=(\partial(F_S^{-1})/\partial f)(F_S(f))$ provides appropriate frequency overlap of the filters.

The AUDlet filters $h_k$ are constructed in the Fourier domain by placing a (symmetric) prototype kernel $w$ at desired center frequencies $f_k$, $k=1,\dots, K$ and impose the bandwidths $B_S(f_k)$ (with an optional factor $\gamma>0$) on the chosen auditory scale $S$, yielding
\begin{equation}\label{eq:AUDlet}
    \hat{h}_k(f)
    = \sqrt{d_k}\cdot w\left(\frac{f-f_k}{\gamma \cdot B_S(f_k)}\right).
\end{equation}
The center frequencies $f_k$ are spaced equidistantly on the image of
$[f_{\min},f_{\max}]$ under $F_S$ and the decimation factors $d_k$ 
are chosen to yield sufficient overlap at low to moderate redundancy, aiming at stability and invertibility. 
If necessary, a low- and a high-pass filter are added to cover the regions $[0,f_{\min})$ and $(f_{\max}, \tfrac{f_s}{2}]$.
The kernels obtained from \eqref{eq:AUDlet}, however, have full length by default. Using them as kernels for a convolutional layer is therefore not practicable, especially if the signal length is very long.
Hence, to efficiently employ such filters in this context, a limitable kernel size is required.

\subsection{Restricted kernel size and a dichotomous auditory scale}

For ISAC, we want to transfer the construction in \eqref{eq:AUDlet} to the time domain and use a (symmetric) prototype kernel with restricted size. To achieve this, a few modifications are required. First, since filters with short kernels have full Fourier-domain support, there is no unique notion of bandwidth that we can match with $B_S$.
We derive a suitable correspondence using a continuous, compactly supported prototype. Let $g\in \mathcal C_c(\RR)$ be supported on $[0, 1]$ with a peak-normalized frequency response ($\Vert \hat{g} \Vert_{\infty}=1$).
By specifying a reference bandwidth\footnote{As a default, we use the $-3$~dB bandwidth of $g$.} $bw_R(g)$ of $g$ we can determine the appropriate dilation that yields a filter kernel with bandwidth $B_S$ when applied to $g$: 
\begin{equation}\label{eq:bw}
    \tilde{B}_{S,g}(f) = \frac{B_S(f)\cdot bw_{R}(g)}{\|g\|^2}.
\end{equation}
The final (discrete) ISAC kernel, associated with center frequency $f$ and bandwidth $B_S(f)$ is obtained by sampling the dilated prototype $g$ at sampling rate $f_s$, and discarding all values that are identically zero. See Eq. \eqref{eq:isac2} in the next section for the precise expression. We denote the resulting kernel size by $T_{S,g}(f)$.
%

As an inherent property of the scaling function, the bandwidths are getting smaller in the lower frequencies, hence, the kernel sizes naturally become larger. If we want to set a maximum kernel size $T_{\max}$ while preserving the bandwidth requirements, we cannot follow the original scale function anymore. As a solution, we linearize it where necessary.
Let $f^*\in [0,\tfrac{f_s}{2}]$ denote the transition frequency where $T_{\max} = T_{S,g}(f^*)$.
We define a modified auditory frequency scale that is the original scale above $f^*$ and its linearization at $f^*$ below (Fig. \ref{fig:scale}). For an auditory scale $S$, the function is given as
\begin{equation*}\label{eq:modaud}
    \tilde{F}_S(f^*;f) =
    \begin{cases}
        F_S(f^*) + F_S'(f^*)(f-f^*)& \text{for } f\in [0, f^*]\\
        F_S(f)& \text{for } f\in (f^*, \frac{f_s}{2}].
    \end{cases}
\end{equation*}
This has the benefit that we can preserve the kernel shape while re-using the bandwidth at $f^*$ (and the corresponding overlap) at all lower frequencies.
Accordingly, we use the following modification of the bandwidth function,
\begin{equation}\label{eq:modbw}
    \tilde{B}_{S,g}(f^*;f) =
    \begin{cases}
        \tilde{B}_{S,g}(f^*)& \text{for } f\in [0, f^*]\\
        \tilde{B}_{S,g}(f)& \text{for } f\in (f^*, \frac{f_s}{2}],
    \end{cases}
\end{equation}
implying that $T_{S,g}(f)=T_{\max}$ for every $f\in [0, f^*]$. If we want to set a minimum kernel size instead, the linearization can be done analogously for the upper frequency region.

\subsection{The ISAC kernels}
Let $g\in \mathcal C_c(\RR)$ be as before, even symmetric around $1/2$
(e.g., a Hann window) and $f^*$ be the transition frequency for the maximal kernel size $T_{\max}$, i.e., $ T_{S,g}(f^*) = T_{\max}$.
For center frequencies $f_k$, $k=1,\dots, K$ obtained from equidistant samples on the image of $[0,\tfrac{f_s}{2}]$ under $\tilde{F}_S(f^*,f)$, the ISAC kernels are given by
\begin{equation}\label{eq:isac2}
    g_k(\ell) = \frac{\sqrt{d_k}}{T_{S,g}(f_k)}\cdot g\left(\frac{\ell}{T_{S,g}(f_k)}\right) \cdot \exp(-2\pi i f_k \ell/L),
\end{equation}
for $\ell = 0, \dots, T_{S,g}(f_k)-1$. To conform with standard implementations of convolutional layers, all kernels with size $T_{S,g}(f)<T_{\max}$ can be zero-padded to size $T_{\max}$ and centered. Recalling that $T_{S,g}(f_k)$ is the kernel size that corresponds to the bandwidth $\tilde{B}_{S,g}(f^*;f_k)$
we see that the above formula is the time domain-version of \eqref{eq:AUDlet}.
The $L_1$-normalization via $T_{S,g}(f_k)^{-1}$ ensures the peak-normalization of the frequency response (provided that $g$ is positive).

Due to the division of the center frequencies into an auditory and a linear part, ISAC can be interpreted as merging an AUDlet filter bank from~\cite{necciari2018AUDlet} with a partial short-time Fourier transform (STFT) with kernel size $T_{\max}$ that covers the low frequency region.
Note that the critical bandwidths of the human auditory sytem are assumed to be approximately constant at low frequencies and increase proportionally to frequency only at higher ranges~\cite{zwicker1961subdivision}. Hence, the frequency scale for ISAC yields a reasonable approximation of auditory scales---provided that $T_{\max}$ is not too small.

Finally, by including $0$ and $\tfrac{f_s}{2}$ as center frequencies, there is no need to add extra low- and high-pass filters, and it allows the user to conveniently choose the number of channels as a second configuration parameter, further facilitating the integration into a neural network.

\begin{table}[t]
    \centering
    \renewcommand{\arraystretch}{1.2} 
    \begin{tabular}{|cc|cccc|}
    \hline
        & & \multicolumn{4}{|c|}{\textbf{kernel size}} \\
        &  & \textbf{8}$^\dagger$ & \textbf{32}  & \textbf{128} & \textbf{512} \\
        \hline
        \multirow{4}{*}{\rotatebox{90}{\textbf{\# channels}}} 
        & \textbf{16}$^\dagger$ & 1.00 & 1.17  & 1.17 & 1.49 \\
        & \textbf{40} & 1.00 & 1.04 & 1.05 & 1.08 \\
        & \textbf{96} & 1.00 & 1.04 & 1.04 & 1.05 \\
        & \textbf{512} & 1.00 & 1.04 & 1.03 & 1.04 \\
    \hline
    \end{tabular}
    \caption{Condition numbers of ISAC filter banks for different kernel sizes and numbers of channels. Decimation factors vary between $1$ and $6$. For the settings with very short filters and few channels (marked with $^\dagger$) we set $\gamma=3$.}
    \label{tab:filters}
\end{table}

\subsection{Stability}
The numerical stability of a filter bank is determined by the condition number of its frame operator matrix~\cite{christensen2002intro}. Let $g_k, d_k$ for $k=1,\dots,K$ denote the kernels and decimation factors for a filter bank with $K$ channels. The condition number is defined as $\kappa = B/A$, where $A$ is the largest, and $B$ the smallest number such that
\begin{equation}
    A \cdot \Vert x \Vert^2 \leq \sum_{k=1}^{K} \Vert (x\ast g_k)\downarrow_{d_k} \Vert^2 \leq B \cdot \Vert x \Vert^2
\end{equation}
for every $x\in \RR^L$. Provided $A>0$ exists, the filter bank is called a frame for $\RR^L$ and the existence of a dual synthesis filter bank with perfect reconstruction is guaranteed.
If $\kappa=1$, the filter bank is said to be a tight frame and possesses convenient properties such as Parseval-type energy preservation and is its own dual filter bank. Figure \ref{fig:analysis} shows that the total PSD of the ISAC filterbank ($K=40$, $T_{\max}=128$), given by $\sum_{i=k}^K \vert \hat{g}_k \vert^2 + \vert \hat{\overline{g}}_k \vert^2$, is nearly constant, which - if decimation is chosen appropriately - provides a good indicator for $\kappa$. However, there are efficient ways to compute $\kappa$ also exactly for uniform decimation~\cite{boelcskei1998filterbanks}. Table \ref{tab:filters} lists the condition numbers of ISAC filter banks for different settings, indicating that they are almost tight in all practically relevant settings.

If the kernels of the filter bank are further optimized, there are different ways to maintain a good condition number~\cite{hasannasab2020proximal, cisse2017parseval}. One simple but effective approach \cite{balazs2024trainable, nenov2024smoothsailing} is to add  $\beta\cdot(\kappa-1)$ as a regularizing term to the loss function, where $\beta>0$ is a hyperparameter that controls the influence of the regularization. If the regularizer kicks in sufficiently, the filter bank can be simultaneously used as synthesis filter bank.

\begin{figure}[t]
    \centering
    \includegraphics[width=\linewidth]{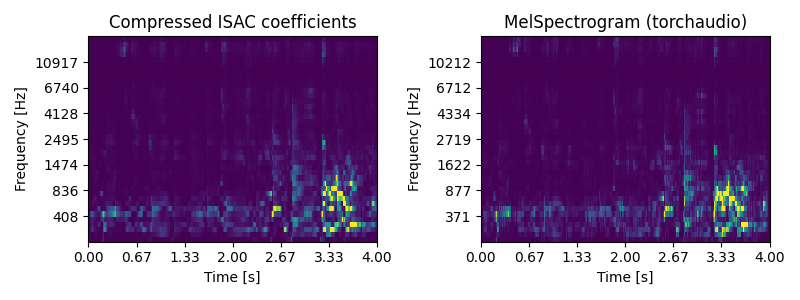}
    \caption{Left: The compressed representation of a glitchy sound based on an ISAC filter bank ($K=40$, $T_{\max}=480$). Right: The mel spectrogram with the same settings. The output sizes match up to a few numbers of time bins.}
    \label{fig:melspec}
\end{figure}

\section{Application}
\subsection{Compression for mel-type spectrograms}

In~\cite{dorfler2020basic} it was shown that a mel spectrogram can be rewritten in terms of a filter bank decomposition using adaptive filters on the mel scale followed by a time-averaging operation after taking the squared modulus of the filter bank coefficients. In short-hand notation,
\begin{equation}
    \vert \text{STFT}(x)\vert^2 \cdot \Lambda = \vert \text{MEL}(x)\vert^2 \ast \phi, 
\end{equation}
where ``$\cdot\; \Lambda$'' shall describe the averaging to obtain the mel spectrogram and ``$\ast\; \phi$'' means channel-wise convolution over time with MEL being the adaptive filter bank.
Although it is in general not known what the time averaging kernel $\phi$ is explicitly, the authors give an example (in the continuous case) where it can only be a Gaussian when a Gaussian window is used in the STFT.
In the finite discrete setting, it is even less clear what $\phi$ should be. A practicable solution that yields a comparable compression as in the mel spectrogram is to use a rectangular window that has the same length as the filters divided by the decimation factor with 50\% overlap. In Figure \ref{fig:melspec} we show the comparison between the proposed compression of a mel-based ISAC filter bank (left) and torchaudio's \texttt{MelSpectrogram} (right), both with $40$ channels and a kernel size of $480$. With the proposed implementation, it is easily possible to make the averaging kernel $\phi$ learnable, yielding an adaptive transform, similar to~\cite{zeghidour2021leaf}.

\subsection{Learn a dual filter bank with the same kernel size}
If a filter bank is a frame, there are infinitely many possible dual frames that provide perfect reconstruction~\cite{christensen2002intro}. Finding a dual frame that exhibits a filter bank structure, yet is not the canonical dual of the filter bank, is challenging, especially if additional properties are desired~\cite{perxxl17} - in our case, restricted kernel support~\cite{stoeva2022gaborcompdual}.
With the integration into an automatic differentiation scheme, we can learn such a synthesis filter bank by minimizing the reconstruction error employing the same kernel size. Thereby, we may find a synthesis filter bank that yields close-to-perfect reconstruction. Since ISAC filter banks are almost tight, the duals are not ``far away'', and we are able to indeed find good approximate dual synthesis filter banks with the kernel size (reconstruction error $\approx 6e-6$). Similar to above, we can promote a flat total PSD via regularization. Such a filter bank is exemplarily depicted in Figure \ref{fig:synthesis}.
In other words, we learn a left-inverse of the analysis operator of the filter bank that is not the Moore-Penrose pseudo inverse (resulting in the canonical dual filter bank which does not have the same kernel size).

This provides a setting that can be efficiently integrated into any neural network-based analysis-synthesis architecture, e.g., for speech enhancement~\cite{haider2024hold}.


\section{Conclusion}
The paper introduces ISAC, an invertible, stable, and perceptually-motivated filter bank with customizable kernel sizes, designed for the needs of modern machine learning.
All transforms are implemented as PyTorch \texttt{nn.module} for a direct integration into PyTorch's automatic differentiation paradigm. With the option \texttt{is\_learnable} the kernels can be set to be learnable parameters of the model. Everything is included in the PYPI package \texttt{hybra} which can be conveniently installed using \texttt{pip install hybra}. The accompanying GitHub project is found under \url{https://github.com/danedane-haider/HybrA-filterbanks}.

\bibliographystyle{abbrv}
\bibliography{references}

\end{document}